\documentclass[aps,prl,twocolumn,groupedaddress]{revtex4-1}

\usepackage{hyperref}
\usepackage{amsmath, mathrsfs, amssymb, wasysym}
\usepackage{graphicx}
\usepackage[caption=false]{subfig}

\begin{document}


\title{Oscillatory athermal quasi-static deformation of a model glass}


\author{Davide Fiocco}
\email{davide.fiocco@epfl.ch}
\affiliation{Institute of Theoretical Physics (ITP), Ecole Polytechnique F\'ed\'erale de Lausanne (EPFL), 1015 Lausanne, Switzerland}
\author{Giuseppe Foffi}
\email{giuseppe.foffi@u-psud.fr}
\affiliation{Institute of Theoretical Physics (ITP), Ecole Polytechnique F\'ed\'erale de Lausanne (EPFL), 1015 Lausanne, Switzerland\\
and\\
Laboratoire de Physique de Solides, UMR 8502, B\^{a}t. 510, Universit\'e Paris-Sud, F-91405 Orsay, France}
\author{Srikanth Sastry}
\email{sastry@tifrh.res.in}
\affiliation{TIFR Centre for Interdisciplinary Sciences, 21 Brundavan Colony, Narsingi, 500075 Hyderabad, India\\
and\\
Jawaharlal Nehru Centre for Advanced Scientific Research, Jakkur Campus, Bangalore 560 064, India.}


\date{\today}


\begin{abstract}

We report computer simulations of oscillatory athermal quasi-static shear deformation of dense amorphous samples of a three dimensional model glass former. 
A dynamical transition is observed as the amplitude of the deformation is varied: for large values of the amplitude the system exhibits diffusive behavior and loss of memory of the initial conditions, whereas localization is observed for small amplitudes. Our results suggest that the same kind of transition found in driven colloidal systems is present in the case of amorphous solids (e.g. metallic glasses). The onset of the transition is shown to be related to the onset of energy dissipation. Shear banding is observed for large system sizes, without, however, affecting qualitative aspects of the transition.

\end{abstract}

\maketitle


Understanding the behavior of materials under mechanical deformation is of primary importance for many contexts. While the deformation behavior of crystals is theoretically well understood, no universally accepted framework exists to rationalize the behavior of mechanically driven amorphous systems, although significant progress has been made in recent years in developing a detailed understanding of how an amorphous solid responds to external stresses \cite{alexander1998amorphous,falk2011deformation, karmakar2010statistical}.  Considerable recent activity has been spurred by an interest in the mechanical behavior of metallic glasses, soft glassy materials, foams and granular packings, and has involved  theoretical, computational and experimental investigations \cite{falk2011deformation,schuh2007mechanical,chen2008mechanical, schall2007structural, chaudhuri2012inhomogeneous,eisenmann2010shear}. Particular interest is understandably focused on the manner in which the response of an amorphous solid changes from nearly elastic response at small applied stress to a state of flow for large applied stress. 

Many computational investigations have employed the approach of studying the zero temperature behavior of amorphous solids under quasi static conditions (using an Athermal Quasi Static or AQS procedure \cite{maloney2006amorphous}). In this procedure, systems are kept in local energy minimum configurations, or inherent structures \cite{stillinger1984packing,stillinger1995topographic} while varying the strain. In previous work on model systems of binary Lennard-Jones particles, it has been shown that upon monotonically increasing the applied shear strain, the inherent structures evolve towards energies corresponding to the limit of high temperatures \cite{utz2000atomistic}. This and related phenomena are referred to as \emph{rejuvenation}, in contrast to the well studied process of \emph{ageing} whereby (typically) a glassy material descends to deeper energy configurations as a function of the waiting time over which it relaxes at a given temperature. In contrast, when a cycle of strain is  applied up to a maximum value which is then reversed \cite{lacks2004energy}, both ageing and rejuvenation are observed, with  small amplitude strains found to reduce the energy of samples (``overage'' them), while larger amplitude strains tend more often to increase the energy (thus ``rejuvenating'' the samples). Initial conditions of the samples in such cases matter: samples with lower initial potential energy are rejuvenated more easily than those with higher energy \cite{lacks2004energy}.  

In a very different context, it has been observed that the nature of rearrangements of colloidal particles in a suspension, at moderate packing fractions, subjected to cyclic shear deformation, depend on the amplitude of strain \cite{pine2005chaos}. For small amplitudes the particles become quiescent after a short transient, whereas for larger amplitudes a finite fraction of the colloidal particles move irreversibly \cite{pine2005chaos,corte2008random}. It has been argued that such a non-equilibrium transition from a quiescent to an active state belongs to the universality class of conserved directed percolation (C-DP) \cite{menon2009universality}.

In this letter we report results of computer simulations for a three dimensional binary Lennard-Jones model, wherein we subject samples to a large number of shear strain cycles (rather than to a single cycle as reported in \cite{lacks2004energy}). Our systems reach, for large enough amplitudes, steady states which are independent of their initial conditions.
Interestingly, initial systems populating different regions of the potential energy landscape converge to the same steady states with an energy that depends only on the maximum strain.  In such cases the systems exhibit diffusive behavior, with a diffusion coefficient that is characteristic of the steady state, depending only on the strain amplitude. For low amplitudes of the strain, samples with different initial conditions fail to converge to a common steady state, and the dynamics is not diffusive, but samples reach instead non-diffusive, quiescent states that strongly retain memory of the initial conditions. Our results thus indicate that sheared amorphous solids undergo the same type of quiescent to active transition as driven colloidal suspensions at moderate density \cite{corte2008random}. 
We find that particle diffusivity is a suitable order parameter for such transition, and suggest that the same diffusion behavior observed in driven colloidal suspensions could occur in completely different cases, namely in fatigue experiments on metallic glasses \cite*{ashby2006metallic, chen2008mechanical,lo2010structural}. 
In addition, we find that the critical amplitude $\gamma_{c}$ at which the transition is observed depends on the size of the samples studied, and is directly related to the dramatic change in the amount of energy dissipated per deformation cycle (the area of the hysteresis curve), rather than to the value of the yield strain of undeformed samples. Finally, we show that large systems subjected to deformation above the critical value present indication of shear banding. Remarkably, while the band can occupy as much as half of the simulation box, the above dynamical transition picture remains unaffected.\\

We consider a binary Lennard-Jones mixture with parameters as in \cite{kob1995testing}. Samples with $N = 500$, $4000$, $32000$ particles are equilibrated at two  different temperatures ($T = 1.0, 0.466$) at reduced density $1.2$. The higher temperature corresponds to the onset of slow dynamics whereas the lower temperature, close to the lowest temperatures studied for this system, corresponds to a supercooled state \cite{sastry1998signatures}. Configurations sampled from the equilibrium trajectory are subjected to energy minimization, using the conjugate-gradient algorithm, to obtain sets of inherent structures typical of the liquid at temperature $T$. The inherent structure energies sampled by the liquid are lower for the lower temperatures \cite{lacks2004energy}. All the simulations were performed using LAMMPS \cite{plimpton1995fast}.\\

\begin{figure}[h]
	\subfloat[][\label{fig:DvsGammaMax}]{\includegraphics[width=0.9\columnwidth]{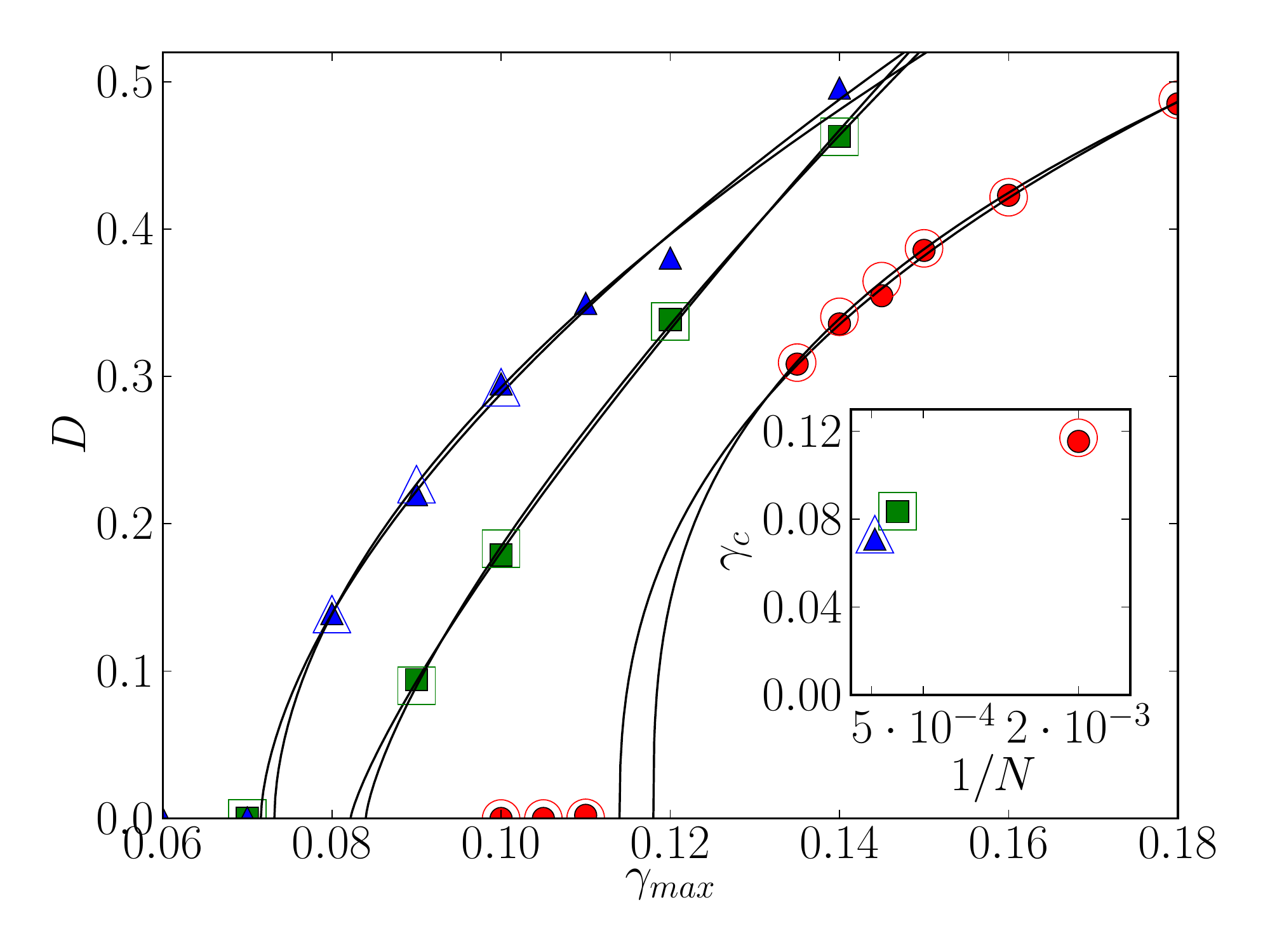}}\\
	\subfloat[][\label{fig:DissipatedEnergy}]{\includegraphics[width=0.9\columnwidth]{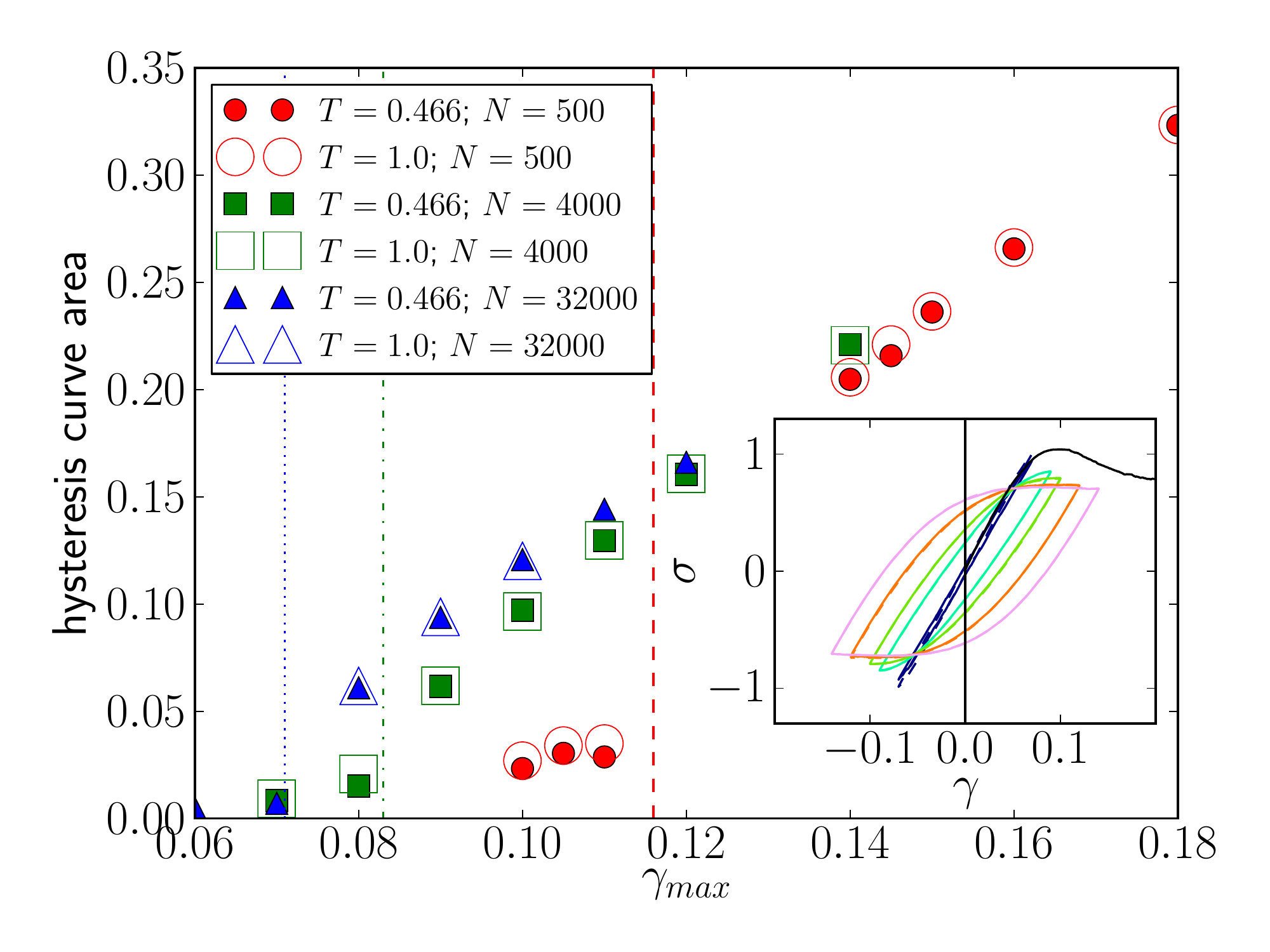}}
	\caption{
	a) Values of diffusivity $D$ obtained by fitting the MSD, for the studied system sizes and temperatures. 
	Superimposed on the data are power law fits (Eq. \ref{eq:PowerLaw}) from which the critical strain ($\gamma_{c}$) values obtained  (for $N = 500, 4000, 3200$ respectively, for ($T = 0.466$, $T = 1.0$)) are: ($0.115, 0.117$), ($0.083, 0.083$) and ($0.070,0.073$). The corresponding exponent values $\beta$ are: ($0.38,0.36$), ($0.75,0.76$), ($0.61,0.54$). The inset shows $\gamma_{c}$ vs. inverse 
	 system size $1/N$. $\gamma_{c}$ extrapolates to a non-zero value as $1/N \to 0$.
	b) Stress-strain curves (shown in the inset for $N = 4000,  T= 0.466$) show hysteresis with finite enclosed area that increases with $\gamma_{max}$
	at high strain amplitude  $\gamma_{max}$. Also shown for reference is the stress-strain curve obtained under uniform shear. The areas enclosed by the hysteresis curves are shown in the main panel as a function of $\gamma_{max}$ for all the different system sizes and temperatures studied. The areas, close to zero at low $\gamma_{max}$, become finite above $\gamma_{c}$ (indicated by vertical lines for each system size).}
\end{figure}

Each of such inherent structures is then subjected to two operations: first, particle positions are  transformed affinely (in this case, we transform the x-coordinates according to $r_x^{'} = r_x + d\gamma r_y$ while keeping $r_y$ and $r_z$ the same)
 so that the shear strain $\gamma$ of the samples is incremented by a small amount $d\gamma$, and Lees-Edwards boundary conditions \cite{lees1972computer} are updated accordingly. 
This has the consequence of changing the original energy landscape in which the inherent structure configuration is not any longer a minimum. Subsequently, energy minimisation is performed using the conjugate gradient algorithm, to obtain the 
inherent structure in the new, \emph{deformed}, energy landscape.  These two operations represent an \emph{Athermal Quasi-Static} (AQS) step \cite{maloney2006amorphous}. If AQS steps are iterated, the sample can be sheared to arbitrary values of strain. The dynamics under such a protocol can be assumed to mimic real shear deformation experiments when the temperature of the system is low enough to be neglected (hence the name athermal) and shear rates are low enough to let the system relax fully before further deformation takes place (so that the deformation can be considered quasi-static) \cite{maloney2006amorphous}. 
In what follows, the shear strain $\gamma$ is changed in steps of $d\gamma$ (= $10^{-3}$, $2 \cdot 10^{-4}$, $2 \cdot 10^{-4}$, respectively for $N = 500$, $4000$, $32000$) from $0$ to a maximum value $\gamma_{max}$ and then to $-\gamma_{max}$ and back to $0$. This periodic operation is the elementary cycle of our dynamics. For practical purposes we define the \emph{accumulated} strain $\gamma_{acc}$ as the sum of the elementary absolute strains applied at all steps: $\gamma_{acc} = \sum_{i} |d\gamma_{i}|.$  The accumulated strain plays the role of time in our analysis. At the end of each cycle, 
particle positions in configurations with  $\gamma = 0$ (zero strain configurations) are stored, so that the mean squared displacement with respect to the initial configuration ($\rm \gamma_{acc} = 0$) can be measured. Energy and all the components of the stress tensor are also recorded at every AQS step (but, except in \ref{fig:DissipatedEnergy}, we report only the zero strain values at the end of each cycle). Up to $40$ samples are subjected to cyclic shear deformation using the protocol above, so that the quantities above can be averaged.

\begin{figure}[h]
	\subfloat[][\label{fig:EnergyvsCumulativeGamma}]{\includegraphics[width=0.9\columnwidth]{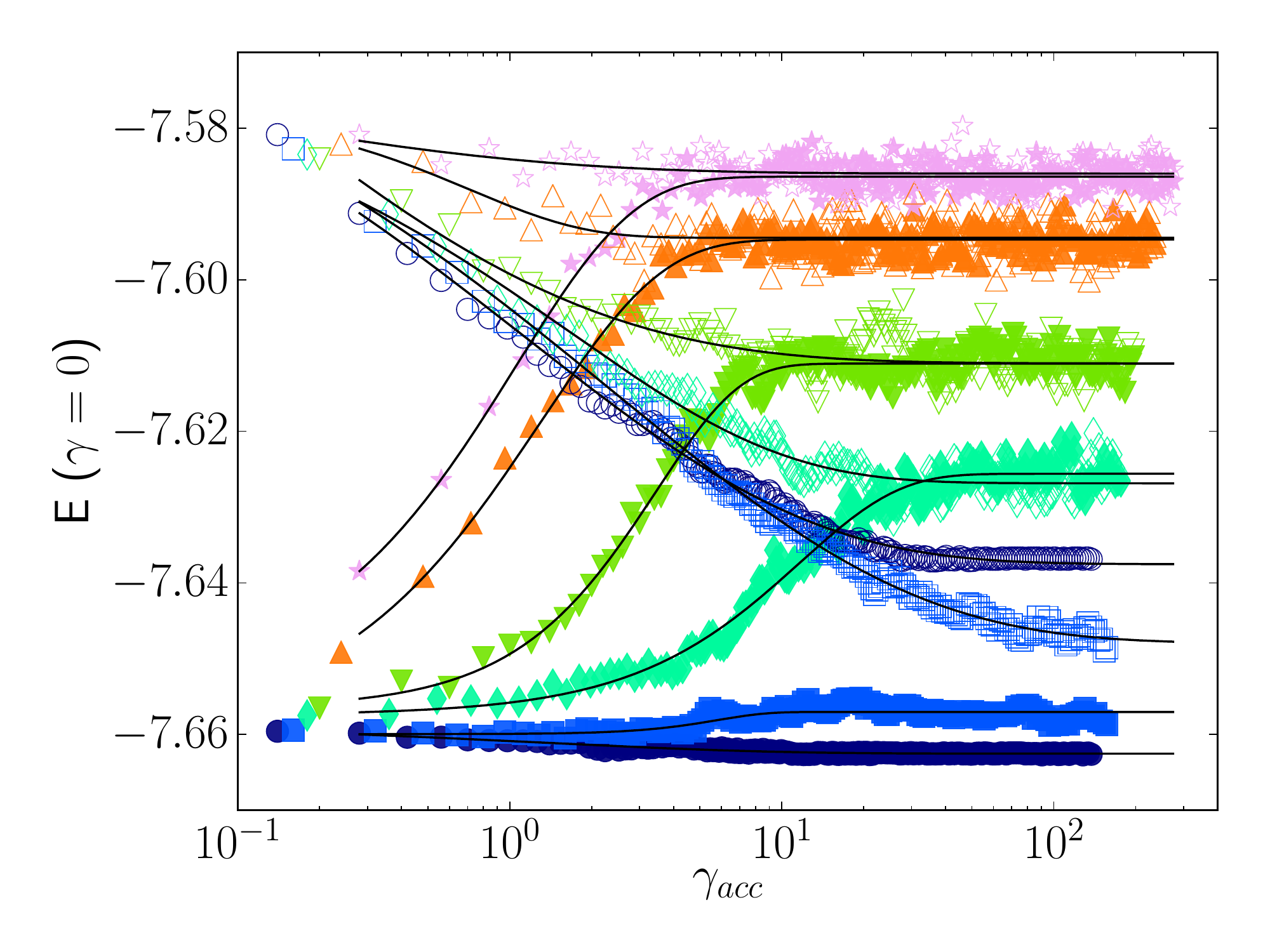}}\\
	\subfloat[][\label{fig:RelaxationPeaks}]{\includegraphics[width=0.9\columnwidth]{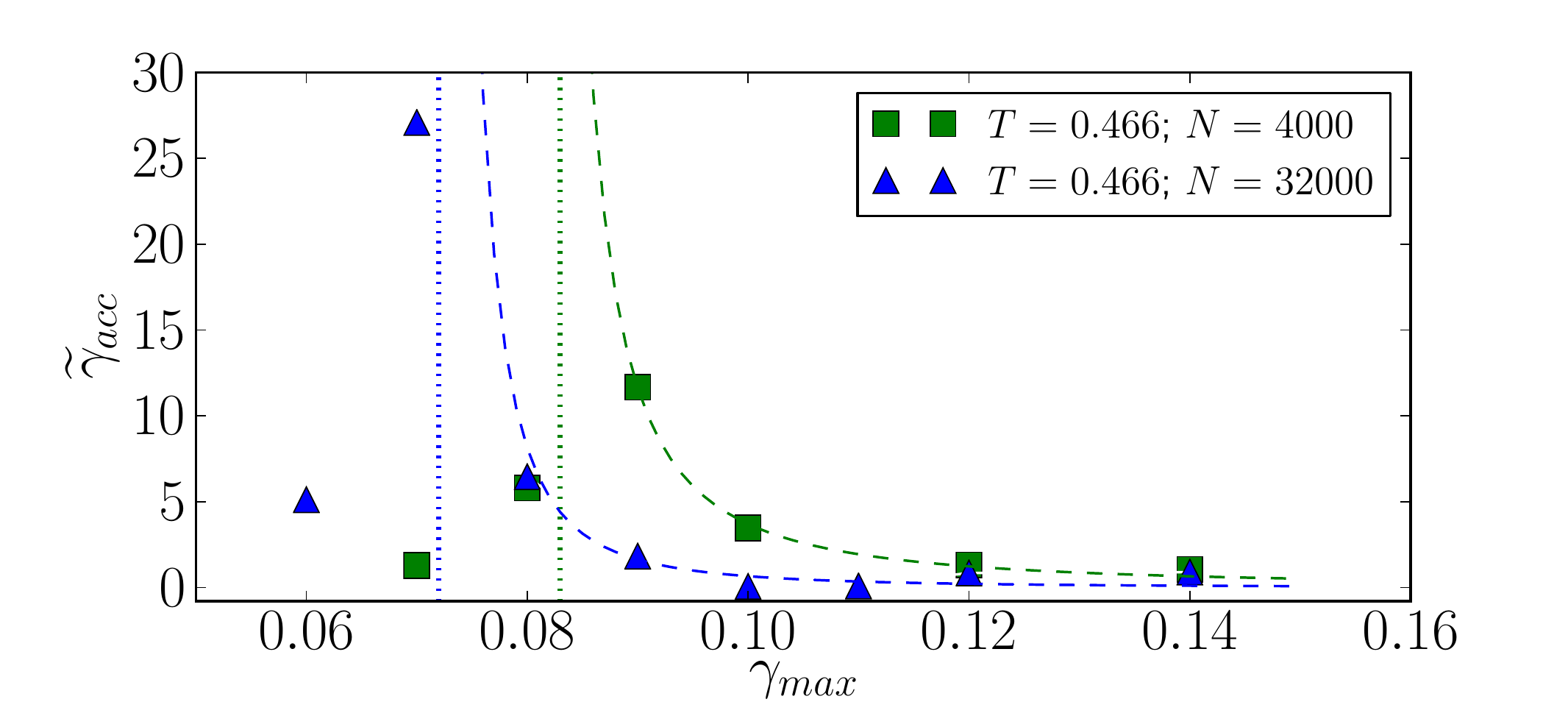}}
	\caption{
	a) Potential energy per particle $E$ for zero strain configurations, for different values of $\gamma_{max}$ (($\bigcirc$):0.07, ($\square$):0.08, ($\Diamond$):0.09 ($\triangledown$):0.1 ($\triangle$):0.12 ($\star$):0.14), averaged over different runs with samples of $N = 4000$ particles at $T = 0.466$ (filled symbols) and $1.0$ (open symbols). The fits are obtained using a stretched exponential model. For the highest values of $\gamma_{max}$, $E$ fluctuates around a plateau after an initial transient, whose value increases with $\gamma_{max}$ and is not dependent on the starting energy (curves with filled and open symbols of the same color/symbol merge), whereas for lower $\gamma_{max}$ memory of initial conditions is retained. For intermediate values our simulations are not long enough to determine which of the two behaviors eventually holds.
	b) Values of characteristic strain $\widetilde\gamma_{acc}$ (akin to relaxation time) obtained from the stretched exponential fits of the energy relaxation in \ref{fig:EnergyvsCumulativeGamma}. The relaxation strain increases as one approaches the transition between the arrested and diffusive regimes. The lines through the data are guides to the eye. Vertical lines indicate critical strain amplitudes $\gamma_c$ described in Fig. \ref{fig:DvsGammaMax}.}
\end{figure}

The average potential energy of the minima visited is presented in Fig.~\ref{fig:EnergyvsCumulativeGamma} as a function of the number of cycles. The behavior of the potential energy depends on the value of $\gamma_{max}$. For low values of $\gamma_{max}$, after a short transient, the energy curves reach a value which depends on the initial $T$. For high enough values of $\gamma_{max}$, the energy \emph{converges} to the same $\gamma_{max}$-dependent value for both the initial values of $T$, and remains steady. We expect this result to hold for other temperatures as well. Whether overaging or rejuvenation will occur, therefore, depends on whether the initial energy of the undeformed inherent structure is lower or higher than the steady state energy. At intermediate values of  $\gamma_{max}$, the convergence of the energy to the asymptotic values requires more cycles than at the extreme  $\gamma_{max}$ values. To characterize this ``slow relaxation'', we fit the transient behavior of the energy data in Fig.~\ref{fig:EnergyvsCumulativeGamma} with a stretched exponential, from which a characteristic relaxation strain $\widetilde\gamma_{acc}$ is extracted.
As can be seen from Fig.~\ref{fig:RelaxationPeaks}, the values of $\widetilde\gamma_{acc}$ become high for intermediate $\gamma_{max}$
and show behavior consistent with a divergence at a critical strain, similarly to what is observed in driven colloidal suspensions \cite{corte2008random}. \\

\begin{figure}[h]
	\includegraphics[width=0.9\columnwidth]{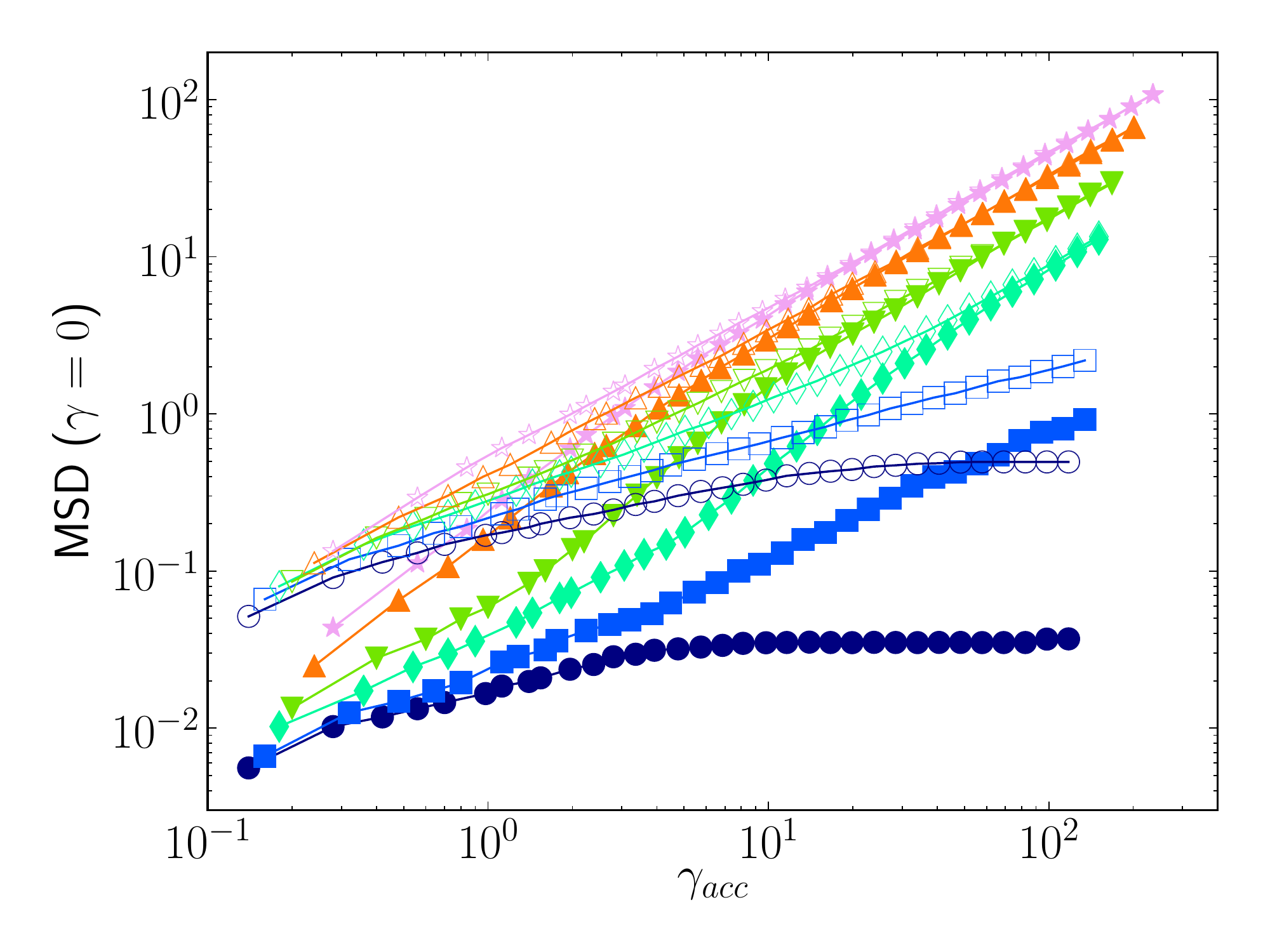}
	\caption{Mean squared displacement for configurations, for different $\gamma_{max}$, averaged over different runs with $N = 4000$ particles at $T = 0.466$ (filled symbols) and $1.0$ (open symbols) (Symbols as in Fig.~\ref{fig:RelaxationPeaks}). Note the transition between an arrested and diffusive regime as $\gamma_{max}$ is increased.\label{fig:MSDvsCumulativeGamma}}
\end{figure}

In order to understand the nature of particle motions that lead to the energy relaxation (or lack thereof) that we observe, we calculate the mean squared displacements (MSD) from the initial configurations, averaged over the samples. The MSD curves shown in Fig.~\ref{fig:MSDvsCumulativeGamma} indicate that for small values of $\gamma_{max}$ the MSD curves saturate after an initial transient, whereas for large $\gamma_{max}$ the particle displacements are \emph{diffusive}, i.e., the MSD depends linearly on the accumulated strain beyond some value of $\gamma_{acc}$. In the case where a steady state is reached, the diffusivity can be extracted by a linear fit of the MSD calculated from an initial point in the steady state. In terms of the relative accumulated strain to such a point, $\gamma_{acc}^{*}$, the MSD is given by
\begin{equation}
	\mathrm{MSD} = D \gamma_{acc}^{*}.
	\label{eq:MSDmodel}
\end{equation}

At intermediate values of $\gamma_{max}$, the system does not reach either an immobile or a diffusive steady state within the number of cycles we perform. The energy in Fig.~\ref{fig:EnergyvsCumulativeGamma} and the MSD data in Fig.~\ref{fig:MSDvsCumulativeGamma}, taken together, suggest that, in this regime, the system undergoes a transition from an  arrested (or quiescent) to diffusive behavior across a critical  $\gamma_{max}$, near which the relaxation of the system becomes very sluggish. 

In Fig.~\ref{fig:DvsGammaMax} the values of $D$ obtained for all the system sizes are plotted against $\gamma_{max}$. For low $\gamma_{max}$,  $D$ is zero and starts to increase rapidly around a critical value  $\gamma_{c}$ which depends on the system size (Fig.~\ref{fig:DvsGammaMax}). The values of $D$ for $\gamma_{max} > \gamma_{c}$ in Fig.~\ref{fig:DvsGammaMax} are reasonably well described by a law of the form (as in \cite{corte2008random})
\begin{equation}
	D = A (\gamma_{max} - \gamma_{c})^{\beta}. 
	\label{eq:PowerLaw}
\end{equation}
The relevant fit values are mentioned in the caption of Fig.~\ref{fig:DvsGammaMax}. Although better analysis is needed to be sure of the values of the exponent $\beta$ and critical strain $\gamma_{c}$, the data shown are clearly consistent with a transition from a regime with zero diffusivity to one with finite diffusivity. In the large strain regime, particles move diffusively and explore configuration space whereas, in the low strain regime, they are localized in configuration space. 
To test the robustness of the observation of a finite $\gamma_{c}$, we  plot  $\gamma_{c}$  for different system sizes in the inset of  Fig.~\ref{fig:DvsGammaMax} against $1/N$. This plot indicates that  $\gamma_{c}$ will be finite for $N \to \infty$.

Interestingly, the transition to the diffusive regime also coincides with the onset of energy dissipation in the system, as can be deduced by considering the stress-strain curves in the steady state, which are shown  in the inset of  Fig.~\ref{fig:DissipatedEnergy}. For small  $\gamma_{max}$, these curves are nearly linear, and do not enclose any significant area, indicating nearly elastic behavior.  
For large strain however, the curves begin to clearly show hysteresis. This effect is common in strain experiments on several materials ranging from metals to bio-materials and it is an indication of a viscoelastic behavior. 
As shown in Fig.~\ref{fig:DissipatedEnergy}, the $\gamma_{max}$ values beyond which dissipation becomes finite agree, to a very good extent, with the critical strain  $\gamma_{c}$ for the corresponding system sizes. These results therefore show that the transition from a localized to a diffusive regime corresponds also to a transition from an nearly-elastic (almost non-dissipative) to a dissipative regime. It is interesting to ask if the critical strain that we observe is related to the yield strain in steady shearing conditions. Although such a possibility is supported by results for the smallest system we study, the yield strain (which we find not to be very sensitive to system size) is higher than the critical strain for the larger ones ($N = 4000, 32000$) (see inset of  Fig.~\ref{fig:DissipatedEnergy}). 

\begin{figure}[h]
	\includegraphics[width=0.6\columnwidth]{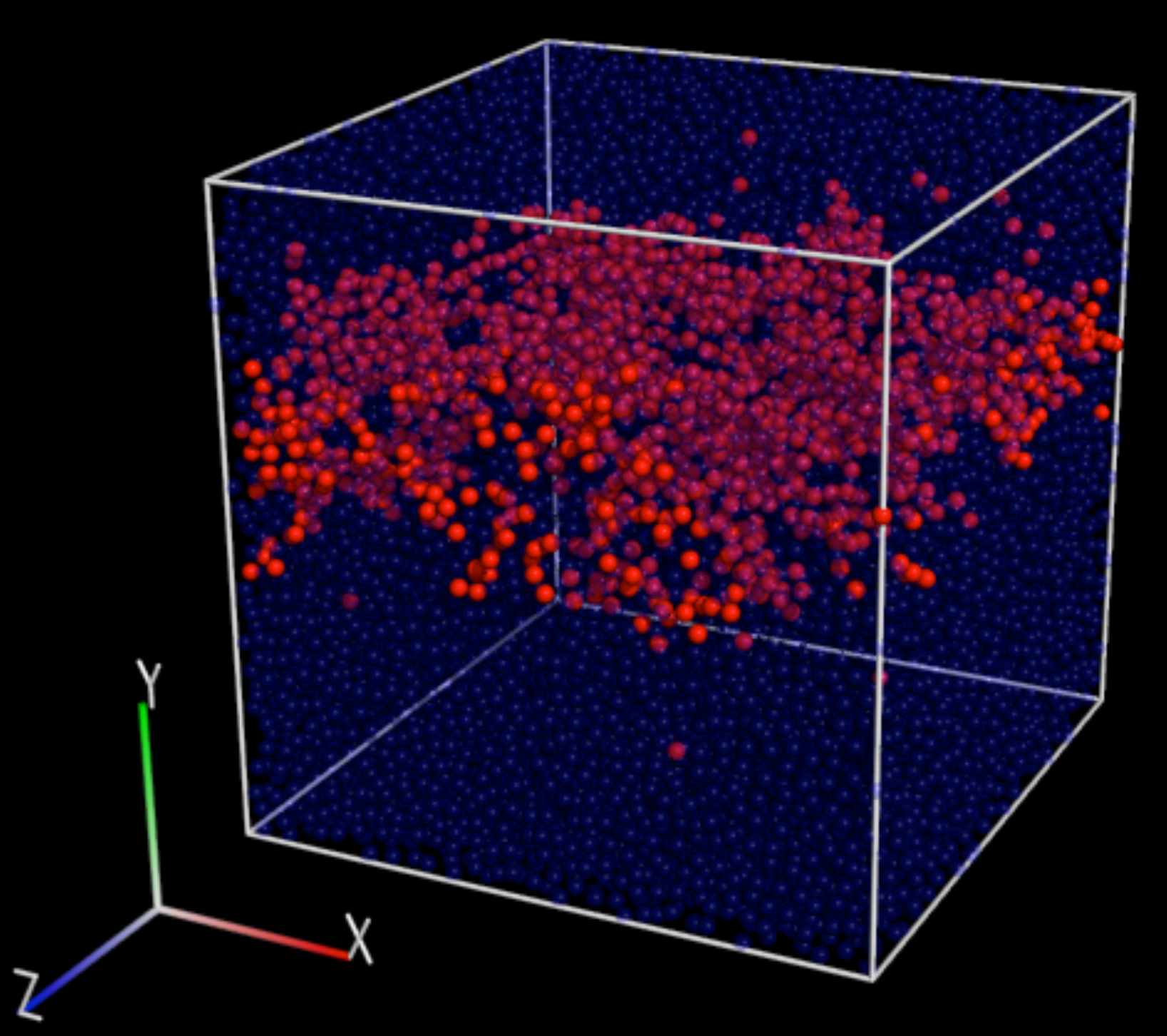}
	\caption{Snapshot of a $N = 32000$ configuration obtained in the steady state for $\gamma_{max} = 0.08$. Particles in red are those that have undergone a scalar displacement $> 0.6\sigma$ when the sample is subject to a full shear deformation cycle on the xy plane. The presence of a band is evident. \label{fig:Band3D}}
\end{figure}

Although our results for different system sizes show qualitatively consistent behavior, at the largest size studied, $N=32000$, a new phenomenon is observed. The system under strain displays clear indications of shear banding. Fig.~\ref{fig:Band3D} shows a snapshot in which  particles that have 
moved  by more than $0.6 \sigma$ in one cycle are highlighted. Such particles are clearly correlated in position, forming a shear band. 
Shear bands are present in all the samples above the critical strain. However, despite the emergence of shear banding, the dynamical transition scenario discussed so far is not affected.

In conclusion, we have shown that athermal oscillatory shear deformation drives, for large enough strains, dense samples of a model glass into steady states in which they explore their energy landscape in a manner dictated by the deformation amplitude  $\gamma_{max}$. Previous observations of overaging and rejuvenation via a single cycle of deformation in a model glass  \cite{lacks2004energy} are thus rationalized as initial steps towards a steady state which is independent of the initial state of the samples and depending only on $\gamma_{max}$. Depending on whether the amplitude of the deformation is below or above a critical strain amplitude  $\gamma_{c}$, these driven amorphous solids can be in an arrested (or localized) state or in a diffusive state which is characterised by hysteresis in the steady state stress-strain curves. The relaxation to the steady state becomes very sluggish for $\gamma_{max}$ values near the critical value.
The transition from a localized to a diffusive steady state we observe is very reminiscent of that observed in experiments on driven systems \cite{corte2008random, bottin2007discontinuous} and reported recently by other authors \cite{regev2013chaos,priezjev2013,Schreck2013} on systems similar to ours. The value of the amplitude $\gamma_{c}$ at which this transition occurs depends on the system size but appears to remain finite in the limit of infinitely large systems. We find that the value of $\gamma_{c}$ has an excellent agreement with the strain value at which hysteresis develops in the stress-strain curves, indicating a transition from a quasi-elastic (non-dissipative)  to a viscoelastic (dissipative) regime.  The critical strain values, however, do not coincide with the yield strain, even though these values are close. Shear banding occurs for our largest system sizes ($N \geq 32000$) in diffusing samples. Interestingly enough, the onset of shear banding at large $N$ doesn't fundamentally affect the dynamical transition. 


\bibliography{FFS}

\end{document}